\documentclass[aps,prb,letterpaper,superscriptaddress,onecolumn,showpacs,floatfix,10pt,notitlepage]{revtex4-1}

\pdfoutput=1

\usepackage{graphicx}
\usepackage{amsmath}
\usepackage{amstext}
\usepackage{amssymb}
\usepackage{amsfonts}
\usepackage{amsbsy} 
\usepackage{dcolumn}
\usepackage{bm}
\usepackage{multirow}

\usepackage{color}
\usepackage[colorlinks,citecolor=blue]{hyperref}

\usepackage[bbgreekl]{mathbbol}
\usepackage{amscd}

\newcommand{\Ref}[1]{Ref.~\onlinecite{#1}}

\newcommand{\om}{\omega}

\newcommand{\bpm}{\begin{pmatrix}}
\newcommand{\epm}{\end{pmatrix}}
\newcommand{\bmm}{\begin{matrix}}
\newcommand{\emm}{\end{matrix}}
\newcommand{\be}{\begin{equation}}
\newcommand{\ee}{\end{equation}}

\usepackage{wrapfig}
\usepackage{subfig}

\begin{document}

\title{Symmetry Fractionalization in Two Dimensional Topological Phases}
\author{Xie Chen}
\affiliation{Department of Physics and Institute for Quantum Information and Matter, California Institute of Technology, Pasadena, CA 91125, USA}

\begin{abstract}
Symmetry fractionalization describes the fascinating phenomena that excitations in a 2D topological system can transform under symmetry in a fractional way. For example in fractional quantum Hall systems, excitations can carry fractional charges while the electrons making up the system have charge one. An important question is to understand what symmetry fractionalization (SF) patterns are possible given different types of topological order and different symmetries. A lot of progress has been made recently in classifying the SF patterns, providing deep insight into the strongly correlated experimental signatures of systems like spin liquids and topological insulators. We review recent developments on this topic. First, it was shown that the SF patterns need to satisfy some simple consistency conditions. More interesting, it was realized that some seemingly consistent SF patterns are actually `anomalous', i.e. they cannot be realized in strictly 2D systems. We review various methods that have been developed to detect such anomalies. Applying such an understanding to 2D spin liquid allows one to enumerate all potentially realizable SF patterns and propose numerical and experimental probing methods to distinguish them. On the other hand, the anomalous SF patterns were shown to exist on the surface of 3D systems and reflect the nontrivial order in the 3D bulk. We review examples of this kind where the bulk states are topological insulators, topological superconductors, or have other symmetry protected topological orders.
\end{abstract}

\maketitle


\section{Introduction}


The discovery of fractional charge in condensed matter systems was unexpected and shocking. In 1982, Tsui, St$\ddot{o}$rmer and Gossard observed that\cite{Tsui1982} when near zero temperature 2D electron gas was subject to a large magnetic field, elementary excitations of the system appear to carry only $1/3$ of the electron charge. This is a puzzling discovery as the energy in the system is definitely too low to actually break up the electrons into pieces. It was then realized that \cite{Laughlin1983,Haldane1983,Halperin1984,Jain1989} the electrons in these fractional quantum Hall (fqH) systems are not broken; instead they participate in a highly nontrivial form of collective motion induced by their strong interaction with each other and the fractional excitations is a manifestation of the underlying `topological' nature of this collective motion.

\begin{figure}[htbp]
\begin{center}
\includegraphics[height=4.0cm]{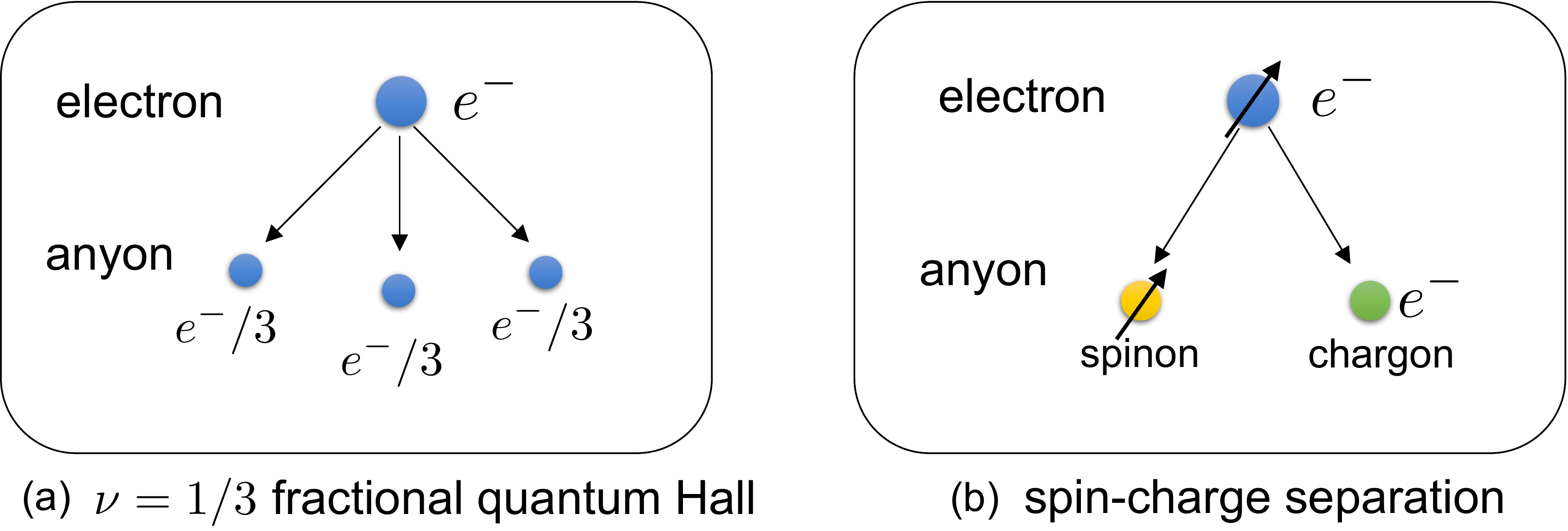}
\end{center}
\caption{Examples of fractionalization. (a) In $\nu=1/3$ fractional quantum Hall systems, one electron fractionalizes into three pieces, each carrying $1/3$ of the electron charge. (b) In spin-charge separated systems, one electron fractionalizes into two pieces: one spinon carrying spin but not charge and one chargon carrying charge but not spin.}
\label{frac}
\end{figure}

Fractionalization can also happen without a large magnetic field. An interesting example is the spin-charge separation, proposed\cite{Haldane1981,Anderson1987,Kivelson1987,Senthil2000} and in some cases observed\cite{Kim1996,Recati2003,Jompol2009,Dalla-Piazza2015} in strongly interacting metals or insulators. It was realized that in these materials, electrons may appear to `split' into spinons and chargons, such that while the electrons carry both spin and charge, the spinons carry only spin but no charge and the chargons carry charge but no spin. Again, this phenomena is possible because the electrons are strongly interacting with each other and fractionalization emerges out of their collective motion. Fig.\ref{frac} illustrates the fractionalization in the $\nu=1/3$ quantum Hall system and the spin-charge separation.

With these inspiring realizations, one is then prompted to ask: what other patterns of fractionalization is possible? Suppose that we start with a many-body system made up of elementary degrees of freedom carrying `integer' quantum numbers / representations of a symmetry, what fractional quantum numbers / symmetry representations can we expect to find in the excitations if local but strong interaction is present in the system?  A lot of progress has been made recently in answering this question in the context of 2D gapped topological phases and this is what we try to summarize in this review. The theoretical development on this topic provides deeper insight into the strongly correlated signatures of topological systems like the spin liquid and topological insulators. This results in new experimental proposals for probing them, which is also discussed in this review. 

The first indication of fractionalization in 2D gapped systems is the existence of `anyons'  -- point excitations that are free to move but have long range braiding or exchange statistics that is different from that of the bosons or fermions making up the systems. The elementary excitation in the $\nu=1/3$ quantum Hall system is such an anyon\cite{Halperin1984}; braiding two such excitations around each other results in a statistical phase factor of $e^{i\pi/3}$. On top of this fractional statistics, if we take into consideration the global charge conservation symmetry of the quantum Hall system, the anyons can further carry fractional quantum numbers of this symmetry. This is also true in the spin-charge separated 2D insulators. The spinons and chargons are anyons of an underlying topological state and have fractional ($-1$) braiding statistics with the vison (another anyon) in the system\cite{Anderson1987,Senthil2000}; moreover, they can carry fractional symmetry representations under global spin and charge conservation symmetry. Therefore, if we want to discuss the full problem of fractionalization, we need to start by asking what set of anyons, hence what topological order, is possible in 2D gapped systems. A lot has been understood in this regard although a full classification has not yet been achieved. But we are not going to focus on this side of the problem. 

Instead, we will assume that the set of anyons in the system (together with all of their fusion and braiding data) is already given. Moreover, we assume the system has certain global symmetry (charge / spin conservation, time reversal, $Z_2$ spin flip etc, lattice translation, lattice rotation, etc). Then we ask, what fractional quantum numbers / symmetry representations can the anyons have? That is, we are interested in how symmetry fractionalizes on the excitations in the system, hence the name `Symmetry Fractionalization'. We are going to use short hand notation `SF' to stand for symmetry fractionalization in this paper.

To address this question, we are going to discuss first, in section \ref{cons}, a simple consistency condition all SF patterns have to satisfy. That is, they have to be consistent with the fusion rules of the anyons. We are going to explain the meaning of this condition and how this leads to a complete --in some sense over-complete-- list of possible SF patterns. It is over complete because, surprisingly, not all consistent SF patterns can be realized in strictly 2D systems. Such SF patterns are called `anomalous'. In section \ref{anomaly}, we discuss examples of anomalous SF patterns and various methods devised to detect such anomalies. With the preparation in section \ref{cons} and \ref{anomaly}, one can then perform a systematic study of SF patterns that can be realized in 2D topological phases with symmetry, the so-called Symmetry Enriched Topological phases. We discuss in section \ref{spliquid} the case of spin liquids, i.e. frustrated spin models with certain topological order and various symmetries like spin rotation, time reversal and lattice symmetries. A list of potentially realization SF patterns can be obtained by excluding the anomalous SF patterns from all consistent ones. It is then interesting to ask which one of them is realized in simple physical models, e.g. the Kagome lattice Heisenberg model. Numerical and experimental probe methods have been proposed which we also review in section \ref{spliquid}. On the other hand, the anomalous SF patterns are not completely impossible. While they cannot be realized in strictly 2D systems, they can appear on the 2D surface of a 3D system. The anomaly in the SF pattern is tightly connected to the nontrivial order in the 3D bulk and we discuss various examples of such connections in section \ref{surface}. Finally, we conclude and discuss open problems in section \ref{discussion}. For pedagogical purpose, we briefly summarize the anyon theory, the notion of symmetry protected topological order and their related gauge theories in the appendix. We refer to these sections in the main text when this background information is necessary for understanding the content.

While we are taking a mostly strongly interacting and lattice regulated perspective in our discussion, we want to mention that mean-field and field theory formalisms can also and have already played an important role in the study of symmetry fractionalization. First, the Projective Symmetry Group\cite{Wen2002,Wang2006} formalism was developed to study spin liquids from a mean-field perspective and proposed the essential idea of using the projective symmetry representations of the anyons to characterize different spin liquids. Not only does it allow classification of spin liquids on various lattice, it also provides a way to write down variational wave functions whose properties can be simulated with Monte Carlo and make connections to realistic Hamiltonians. Secondly, field theory analysis has lead to important results in the study of symmetry fractionalization. For example, it has been shown how to construct, classify, and gauge a large class of symmetry enriched topological phases using the Chern-Simons description\cite{Levin2009,Lu2016SET,Hung2013,Hung2013_Kmatrix}, derive surface SF pattern from bulk topological term\cite{Vishwanath2013, Bi2015}, and detect anomalies through gauge non-invariance of the action\cite{Kapustin2014,Kapustin2014arxiv} .

\section{Consistency Condition}
\label{cons}

To understand the consistency condition the SF patterns have to satisfy, let's go back to the $\nu=1/3$ fqH example. The elementary anyonic excitation in the system can be thought of as $1/3$ of an electron. That is, if we put three of them together, they are equivalent to a single electron excitation. Therefore, the amount of charge they each carry has to be $1/3$ of an electron charge. If they were to carry any other different fractional amount of charge (like $1/2$ or $1/5$), it will not be consistent. Similarly, in the case of spin-charge separation, the combination of a spinon and a chargon is equivalent to a single electron. This is consistent with the SF pattern on the spinons and chargons, as the sum of their charge / spin quantum numbers equals that of a single electron. Such a consistency condition can be readily generalized to all 2D anyon theories, as discussed in \Ref{Wen2002,Kitaev2006,Essin2013,Barkeshli2014arxiv}. For a brief review of 2D anyon theory, see Appendix A. To present this consistency condition, we need to introduce the concept of projective symmetry action on the anyons.

\begin{figure}[htbp]
\begin{center}
\includegraphics[height=3.5cm]{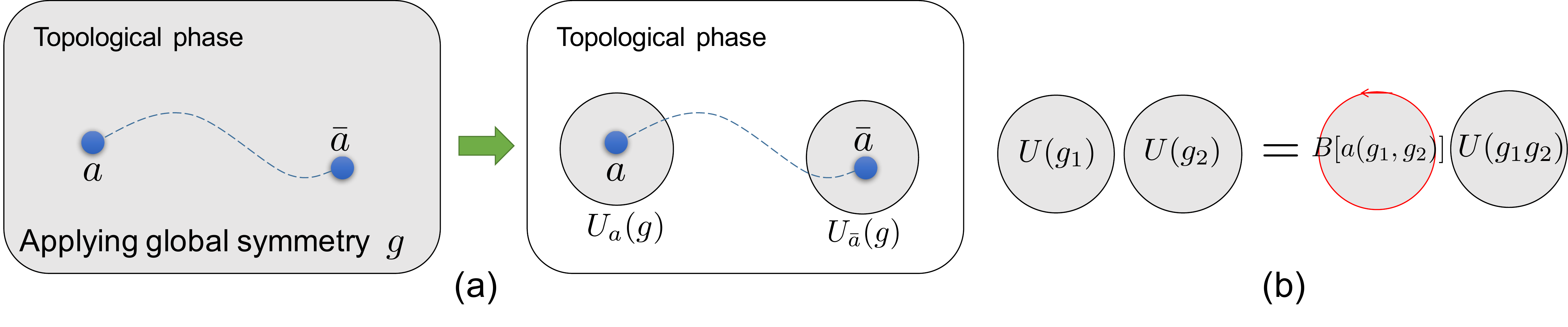}
\end{center}
\caption{(a) Global symmetry action on a state with isolated anyons can be reduced to symmetry actions on each anyon individually, (b) which form a `projective' representation of the group. 
}
\label{proj}
\end{figure}

 Consider a 2D topological phase containing anyon types $a$, $b$, $c$ etc. and with global symmetry of group $G$. The ground state, which has a finite energy gap to all the excited states, is invariant under global symmetry action. Now imagine creating a pair of anyons $a$ and $\bar{a}$ and moving them to distant locations, as shown in Fig.\ref{proj}. If we apply global symmetry $g\in G$ to the system, everywhere away from the anyons the system remains invariant as the state is the same as in the ground state, while near the anyons the system may be transformed in a nontrivial way. If the symmetry $g$ does not change anyon types (which is the case we focus on in this review), then the transformation is equivalent to some local unitaries $U_a(g)$ and $U_{\bar{a}}(g)$ near the anyons, as shown in Fig.\ref{proj}. That is to say, global symmetry action on a state with isolated anyons can be reduced to symmetry actions on each anyon individually. 

These local symmetry transformations are special in that they only have to satisfy the group relation up to a phase. That is,
\be
U_a(g_1)U_a(g_2) = \omega_a(g_1,g_2) U_a(g_1g_2)
\ee
where $\omega_a(g_1,g_2)$ is a phase factor that is not necessarily $1$. $U_a(g)$ is said to form a `projective' representation of group $G$.This is not contradicting to the global symmetry of the system as long as the phase factor on $\bar{a}$ -- $\omega_{\bar{a}}(g_1,g_2)$ -- cancels with $\omega_a(g_1,g_2)$. On the other hand, on the fundamental degrees of freedom making up the system (e.g. electrons) or any excitation created through local operation, $\omega$ has to be $1$ and $U(g)$ is said to form a `linear' representation on these non-fractionalized objects.
For example, if we apply two consecutive $\pi$ rotations in the $U(1)$ symmetry group of charge conservation to the $1/3$ electron excitation in the $\nu=1/3$ fqH, we get
\be
U_{1/3}(\pi)U_{1/3}(\pi) = e^{i2\pi/3}U_{1/3}(0)
\ee
Note that as two excitations related to each other through local operations are considered to be of the same anyon type (see Appendix A), two projective representations differing by a linear representation correspond to the same SF pattern, as described by the same $\omega$. In the fqH case, this tells us that only the fractional part of the charge carried by the anyon matters.

The associativity condition that $\left(U_a(g_1)U_a(g_2)\right)U_a(g_3)=U_a(g_1)\left(U_a(g_2)U_a(g_3)\right)$ leads to the requirement that
\begin{align}
 \om_a^{s(g_1)}(g_2,g_3)\om_a(g_1,g_2g_3)&=
 \om_a(g_1,g_2)\om_a(g_1g_2,g_3),
 \label{2cocycle_om}
\end{align}
which needs to be satisfied by any $\omega$ describing a projective symmetry action. Moreover we have the freedom to change the definition of each $U_a(g)$ by an arbitrary phase factor $\mu_a(g)$. Therefore, $\omega_a(g_1,g_2)$ and $\omega'_a(g_1,g_2)$ related as
\be
\omega'_a(g_1,g_2) = \frac{\mu_a(g_1)\mu_a^{s(g_1)}(g_2)}{\mu_a(g_1g_2)}\omega_a(g_1,g_2)
\label{2coboun_om}
\ee
are considered equivalent. Here $s(g_1)=1$ if $g_1$ is unitary and $s(g_1)=-1$ if $g_1$ is anti-unitary (time reversal). 
Eq.\ref{2cocycle_om} and \ref{2coboun_om} defines the equivalence classes of projective representations, which is an essential concept in the consistency condition of SF patterns discussed below.

The consistency condition of SF patterns then states that: \emph{if $c$ is (one of) the fusion product of $a$ and $b$, then the projective representation carried by $c$ should be equivalent to the tensor product of that carried by $a$ and $b$.}  That is, if the anyons obey fusion rules $a \times b = \sum_c N^c_{ab} c$,  then when $N^c_{ab} \neq 0$,
\be
\om_a (g_1,g_2) \om_b(g_1,g_2) \sim \om_c(g_1,g_2)
\ee
Here we are using `$\sim$' instead of `$=$' because the $\om$'s only have to be equivalent as describe in Eq.\ref{2coboun_om}. In the $\nu=1/3$ fqH case, this condition is simply saying that the fractional part of the charge carried by three elementary excitations should sum to one.

What is the physical meaning of this projective phase factor $\om$? It turns out that $\om$ relates the local action of symmetry to anyon braiding statistics. Consider again the $\nu=1/3$ fqH example. Applying two consecutive $\pi$ rotations to a region $D$ $D$ is equivalent to doing nothing up to a phase factor. The phase factor is $e^{i2\pi n/3}$ if we have $n$ elementary anyons in the region $D$. On the other hand, this is exactly the phase factor induced by braiding an elementary anyon around this region $D$, as shown in Fig.\ref{proj}. Therefore, we have the following relation
\be
U(\pi)U(\pi)  = B[a_{1/3}] U(0)
\ee
where $B[a_{1/3}]$ denotes the braiding of the $1/3$ electron around region $D$. Note that this relation holds no matter what the anyon content is in the region $D$. In general, we have
\be
U(g_1)U(g_2) = B[a(g_1,g_2)] U(g_1g_2)
\label{proj_fs}
\ee
where $a(g_1,g_2)$ takes value in the set of abelian anyons of the system which we denote as $A$. $B[a(g_1,g_2)]$ denotes the braiding of $a(g_1,g_2)$ around region $D$, such that $\omega_b(g_1,g_2)$ is equal to the braiding statistics between $b$ and $a(g_1,g_2)$. (For more careful explanation on this relation see \Ref{Barkeshli2014arxiv}) It is easy to check that the set of projective representations carried by the anyons (as described by $\omega$) generated in this way automatically satisfy the consistency condition described above, due to the additivity of braiding statistics with abelian anyons (see Appendix A). Here we are discussing the local symmetry action $U(g)$ in an abstract way. In section \ref{gauging} we are going to discuss explicitly how to find the operator $U(g)$ for a gapped symmetric state and why they satisfy the relation in Eq. \ref{proj_fs}.

Equation \ref{proj_fs} hence describes the SF patterns on all the anyons in the topological phase together. We can interpret the $U(g)$'s in Eq.\ref{proj_fs} as a projective representation of the symmetry group $G$, but not with coefficient in phase factors as in the case of $U_a(g)$, but rather with coefficient in $A$, the set of abelian anyons of the system. The conditions the abelian anyon coefficient $a(g_1,g_2)$ has to satisfy is completely analogous to that for the phase factor coefficient $\omega_a(g_1,g_2)$. First, associativity requires that
\be
a(g_2,g_3)\times a(g_1,g_2g_3)=
 a(g_1,g_2)\times a(g_1g_2,g_3),
 \label{2cocycle_al}
\ee
Here `$\times$' denotes fusion of abelian anyons. Two sets of anyon coefficient are equivalent if they can be related as
\be
a'(g_1,g_2) = b(g_1) \times b(g_2) \times \bar{b}(g_1g_2) \times a(g_1,g_2)
\label{2coboun_al}
\ee
where for any choice of $b(g) \in A$. This relation comes from redefining the local symmetry action $U(g)$ by braiding $b(g)$ around the region. Using Eq. \ref{2cocycle_al} and \ref{2coboun_al}, we can then find all possible and distinct SF patterns. Mathematically speaking, the set of SF patterns is classified by the second cohomology group of $G$ with coefficient in abelian anyons $A$, denoted as $H^2(G,A)$. This result can be generalized to the situation where anyon types are permuted under symmetry transformation, as discussed in \Ref{Barkeshli2014arxiv,Tarantino2015arxiv}.

Note that while the above discussion is based on internal symmetries like spin rotation or time reversal, it can be applied to spatial symmetries as well\cite{Wen2002,Essin2013}. Consider, for example, the case of reflection symmetry $R$. Imagine creating a pair of anyons $a$ and $\bar{a}$. If $a$ is the same as $\bar{a}$, this can be done in a way that preserves reflection symmetry. One can then decompose the action of reflection symmetry as spatially exchanging the position of the two $a$'s, together with some local unitaries $U_{a}(R)$ around each of the $a$'s. $U_{a}(R)$ then effectively acts as a internal $Z_2$ symmetry on $a$ and its possible fractionalization patterns follow from that of $Z_2$.

Let's see apply the above classification conclusion to the case of a $Z_2$ spin liquid with time reversal symmetry. A spin liquid with $Z_2$ topological order, as reviewed in Appendix A, contains four abeilan anyons $\{\mathbb{1},e,m,f\}$. If the system has global time reversal symmetry, which satisfies $\mathcal{T}\mathcal{T}=I$, locally the symmetry action may act as
\be
U(\mathcal{T})U^*(\mathcal{T}) = B[a(\mathcal{T},\mathcal{T})], \ \ a(\mathcal{T},\mathcal{T}) = \mathbb{1}  \ \text{or} \ e  \ \text{or} \ m  \ \text{or} \ f
\ee
Note that the second $U(\mathcal{T})$ is complex conjugated because the first symmetry action is anti-unitary\cite{Levin2009}. These four cases exhaust all possible SF patterns in the $Z_2$ spin liquid and they are not related to each other through Eq.\ref{2coboun_al}. What does this mean physically? Suppose that $U(\mathcal{T})U^*(\mathcal{T})=e$. This is saying that applying time reversal twice on $m$ or $f$ gives $-1$, while applying time reversal twice on $\mathbb{1}$ or $e$ gives $+1$. That is, the anyons $m$ and $f$ transform as a Kramer doublet under time reversal and anyon $e$ (and the vacuum $\mathbb{1}$) transform as a singlet.

One can apply this procedure to spin liquids with all kinds of symmetries and completely list all consistent SF patterns. For example, in \Ref{Essin2013}, this procedure has been carried out for $Z_2$ spin liquid with square lattice space group, time reversal and $SO(3)$ spin rotation symmetries, where $2^{21}$ different types of SF patterns where identified. 

So there is a huge number of SF patterns that satisfy the consistency condition. Can they all be realized in local two dimensional systems?

\section{Anomaly Detection}
\label{anomaly}

One important realization about symmetry fractionalization is that, among all the consistent SF patterns, not every one can be realized in strictly 2D systems. The ones that cannot are said to be `anomalous'. At first sight this may seem surprising, as the consistent SF patterns (for example the four possibilities for time reversal in $Z_2$ spin liquid) look very much well-behaved and there seems to be no particular reason to suspect that one is more exotic than another. In fact, one has to look very carefully to see the distinction. In this section, we first give some examples of how such anomalies can occur, and then introduce a number of methods that have been developed to detect anomalies in this situation. We wouldn't be able to explain every method in detail, but we will give a careful description of the powerful idea of `gauging', which is at the core of most of the anomaly detection methods.

\subsection{Anomalous SF Pattern: Examples}
\label{anom:example}


Consider a system with $Z_2$ topological order (see Appendix A for review) and charge conservation symmetry. The SF pattern is given by the fractional charge carried by the anyons $e$, $m$ and $f$. Note that we are always talking about the charge amount mod $1$, as the integer part of the charge does not matter. One consistent SF pattern is that $e$ and $m$ carry charge $1/2$ while $f$ carries charge $0$. In terms of the projective representation formed by local symmetry operations as given in Eq.\ref{proj_fs}, we have
\be
a(\theta, 2\pi-\theta) = f, \ \ a(\theta_1,\theta_2) = \mathbb{1} \text{\ otherwise\ }
\ee
Following the terminology of \Ref{Wang2013}, we call this SF pattern $eCmC$. This SF pattern on its own CAN be realized in strictly 2D systems, as can be seen from the following Chern-Simons field theory description:
\be
\mathcal{L} = \frac{2}{4\pi} \epsilon^{\lambda\mu\nu} \left(a^1_{\lambda}\partial_{\mu}a^2_{\nu} + a^2_{\lambda}\partial_{\mu}a^1_{\nu}\right)-\frac{e}{2\pi} \epsilon^{\lambda\mu\nu} \left(A_{\lambda}\partial_{\mu}a^1_{\nu} + A_{\lambda}\partial_{\mu}a^2_{\nu}\right)
\label{Z2CS}
\ee
where $a^1$ and $a^2$ are internal gauge fields and $A$ is external electro-magnetic field. A detailed review of the Chern-Simons formalism can be found for example in \Ref{Levin2011,Lu2016SET} and from there it is straight forward to see that the topological order described by Eq.\ref{Z2CS} is indeed that of a $Z_2$ topological order and the $e$ and $m$ anyons carry half charge.

\begin{wrapfigure}{r}{0.5\textwidth}
  \vspace{-15pt}
  \begin{center}
    \includegraphics[width=0.240\textwidth]{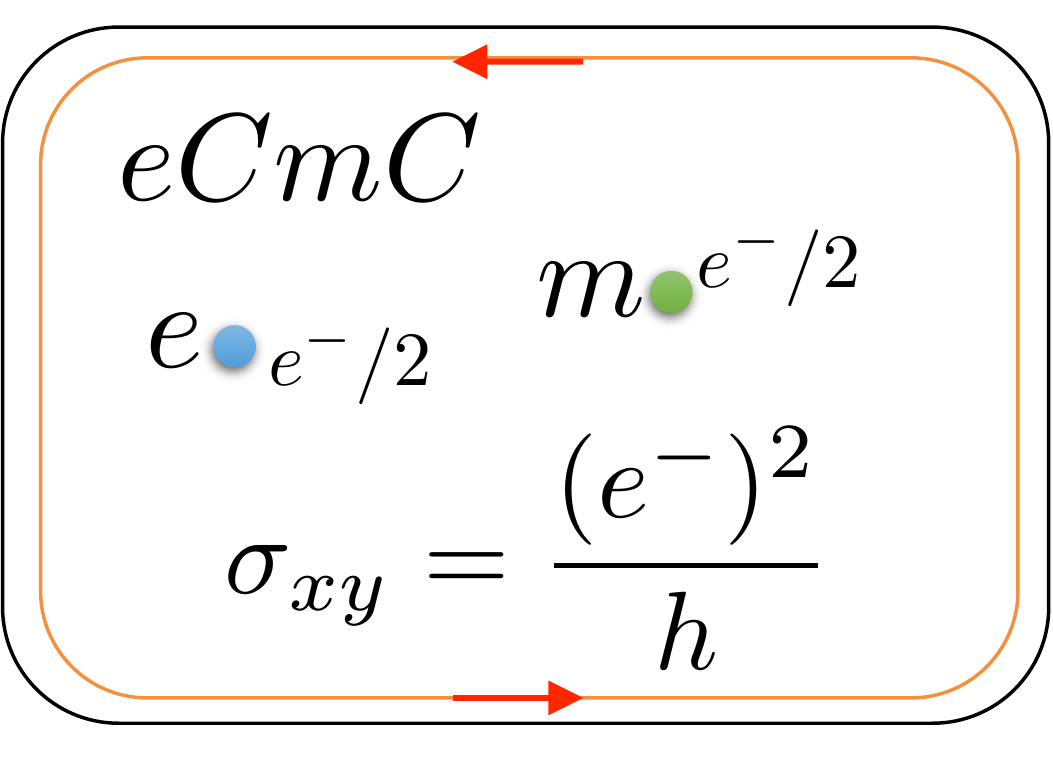}
  \end{center}
  \vspace{-10pt}
  \caption{The $eCmC$ SF pattern: in 2D $Z_2$ topological state if both the $e$ and $m$ anyon carry half charge under charge conservation symmetry, the system has nonzero Hall conductance, hence explicitly breaks time reversal symmetry.}
  \vspace{-15pt}
    \label{eCmC}
\end{wrapfigure}

However, if in addition to charge conservation symmetry time reversal symmetry is also required, this SF pattern becomes anomalous\cite{Vishwanath2013,Wang2013}. To see this, suppose that we can realize the theory described in Eq.\ref{Z2CS} in 2D with both charge conservation and time reversal symmetry. Imagine puting it on a disc with boundary, as shown in Fig.\ref{eCmC}. Following \Ref{Levin2011,Lu2016SET}, we can see that the system has a Hall conductivity $\sigma_{xy} = \frac{(e^-)^2}{h}$. That is to say, there is a chiral edge mode going around the boundary, which explicitly violates time reversal symmetry. Therefore, we get a contradiction and the $eCmC$ SF pattern is not possible in 2D systems with time reversal symmetry. That is, with time reversal symmetry $eCmC$ is an anomalous SF pattern, even though it is consistent.

The $eCmC$ SF pattern provides an example where anomalies under time reversal symmetry can be detected by looking for chiral edge modes on the boundary. Anomalies with unitary symmetries, on the other hand, can only be revealed with more sophisticated methods. In the following sections, we are going to discuss various anomaly detection methods which allow us to see the anomaly in
\begin{enumerate}
\item $eCmT$ -- $Z_2$ topological order with $U(1)$ spin conservation and time reversal symmetry (the two symmetries commute); $e$ carries half spin, $m$ carries integer spin; $e$ is a time reversal singlet, $m$ is a time reversal doublet.
\item $Z_2$ topological order with $Z_2\times Z_2$ symmetry; $e$ transforms as $i\sigma_z$, $\sigma_x$ under the two $Z_2$'s; $m$ transforms as $I$, $i\sigma_y$ under the two $Z_2$'s. ($\sigma_x$, $\sigma_y$, $\sigma_z$ are Pauli matrices.)
\item Projective semion -- chiral semion topological order (see Appendix A for review) with $Z_2\times Z_2$ symmetry; the semion transforms as $\sigma_x$, $\sigma_z$ under the two $Z_2$'s.
\end{enumerate}

Before we discuss particular anomaly detection methods, we introduce the idea of `gauging', which is essential for the following discussions.

\subsection{Gauging the global symmetry}
\label{gauging}


If the system has a global on-site unitary symmetry $G$ acting as a tensor product over all lattice sites $\prod_k M_k(g)$, we can gauge it, as described in \Ref{Swingle2014,Levin2012,Barkeshli2014arxiv}. For a lattice Hamiltonian $H=\sum_i h_i$, this corresponds to a concrete procedure of writing down a new Hamiltonian based on the old one. For simplicity, we describe the procedure when $G=Z_n$. For other groups $G$, a similar procedure can be applied.
\begin{enumerate}
\item WLOG, suppose that the degrees of freedom ($\rho$) in $H$ lives on the vertices of a 2D lattice and on each $\rho$ the generator of the symmetry acts as $M=\rho^x$, such that $(\rho^x)^n=I$.  Introduce new degrees of freedom, the `gauge field' $\tau$, to each edge of the lattice. The `gauge field' are bosonic degrees of freedom and has basis states $|g\rangle, g=0,...,n-1$. Define operators $\tau^x |g\rangle = |g+1 \text{\ mod\ } n\rangle$ and $\tau^z|g\rangle = e^{i2\pi g/n}|g\rangle$.
\item Assign a direction to each edge. Define a local version of the symmetry such that the symmetry action of the group generator at site $k$ acts not only on $\rho_k$ as $\rho_k^x$, but also on all the edges connected to site $k$ as $\tau^x_{kk'}$ if the edge points away from $k$ and $\left(\tau^x_{kk'}\right)^{-1}$ if the edge points towards $k$.
\item Modify each $h_i$ term in $H$ into $h'_i$ by adding $\tau$ operators on the surrounding edges such that each term is invariant under all local symmetry transformations. For example, the Ising model $H=\sum_{<vv'>} \sigma^z_v\sigma^z_{v'}$ with nearest neighbor coupling has a global $Z_2$ symmetry $\prod_v \sigma^x_v$. By introducing gauge field $\tau_{vv'}$ and modifying the Hamiltonian terms as $\sigma^z_v\tau^z_{vv'}\sigma^z_{v'}$, we can make each term invariant under local symmetry transformation $\sigma^x_v\prod_{<vv'>} \tau^x_{vv'}$.
\item For each plaquette $p$ in the lattice, make a product of $\tau^z$ on all the edges around the plaquette $B_p=\prod_{e\in \partial p} \tau^z_e$ and add it (and its Hermitian conjugate if not Hermitian) to the Hamiltonian with a minus sign. This term enforces zero gauge flux in the ground state.
\item For each lattice site $v$ in the lattice, define operator $A_v=\rho^x_v \prod_{<vv'>}\left(\tau^x_{vv'}\right)^{s_{vv'}}$ ($s_{vv'}=\pm 1$ depending on the orientation of $vv'$) and add it (and its Hermitian conjugate if not Hermitian) to the Hamiltonian with a minus sign. This term imposes the Gauss's law on the gauged model.
\end{enumerate}

After this procedure, we obtain a new Hamiltonian on the expanded Hilbert space of the form
\be
H' = \sum_i h'_i -\sum_p B_p - \sum_v A_v
\ee
All the $B_p$ and $A_v$ terms commute and they commute with the $h'_i$ terms. Therefore, if originally $H$ is gapped, then $H'$ is also gapped. The anyons in $H$ becomes anyons in $H'$. Moreover, new types of anyons emerge in $H'$ which correspond to flux and charge excitations of the gauge field. Therefore, after this gauging procedure, the original topological order gets expanded into a more complicated topological order.

How is this related to anomaly detection of SF patterns? If we know that certain SF pattern can be realized by a 2D lattice model, we can apply this procedure and find the expanded topological order starting from the original SF pattern. But of course in many case, we do not know if the assumption is true. In fact, this is exactly what we are trying to determine. What can help us, is a close connection between the original SF pattern and the expanded topological order if the gauging process can be carried through. Partial information about the expanded topological order, including its anyon types, part of their fusion rules and braiding statistics, can be determined from the SF pattern alone, without specific knowledge about its lattice realization. From this information one can determine if the gauging process can be carried through, resulting in a consistent expanded topological order. In some cases, however, we can see that there are inconsistencies in the gauging process, preventing the gauging process from completing. Such an obstruction to gauging indicates the existence of anomaly in the original SF pattern. This is the underlying logic behind the methods we describe in the following sections, where different methods provide different ways to reveal inconsistencies in the gauging process. 

One important piece of information about the expanded topological order that can be extracted from the SF pattern is the fusion rules of gauge fluxes. Gauge fluxes are excitations in the gauged theory that violate the $B_p$ term (the zero flux rule), but to determine their fusion rules it suffices to think in terms of the ungauged Hamiltonian $H$\cite{Barkeshli2014arxiv,Tarantino2015arxiv}. In $H$, gauge fluxes show up as symmetry fluxes at the end point of symmetry defect lines and, upon gauging, they become deconfined anyonic gauge flux excitations of $H'$.  By examining the symmetry fluxes $\Omega_g$ of $H$, we will show that they obey the `projective' fusion rule 
\be
\Omega_{g_1} \times \Omega_{g_2} = a_{g_1,g_2}\Omega_{g_1g_2}
\label{proj_flux}
\ee
which descends into the fusion rules of the gauge fluxes upon gauging. As we are going to see, this `projective' fusion rule follows from the local projective symmetry action of a SF pattern given in Eq.\ref{proj_fs}.

\begin{figure}[htbp]
\begin{center}
\includegraphics[height=4.0cm]{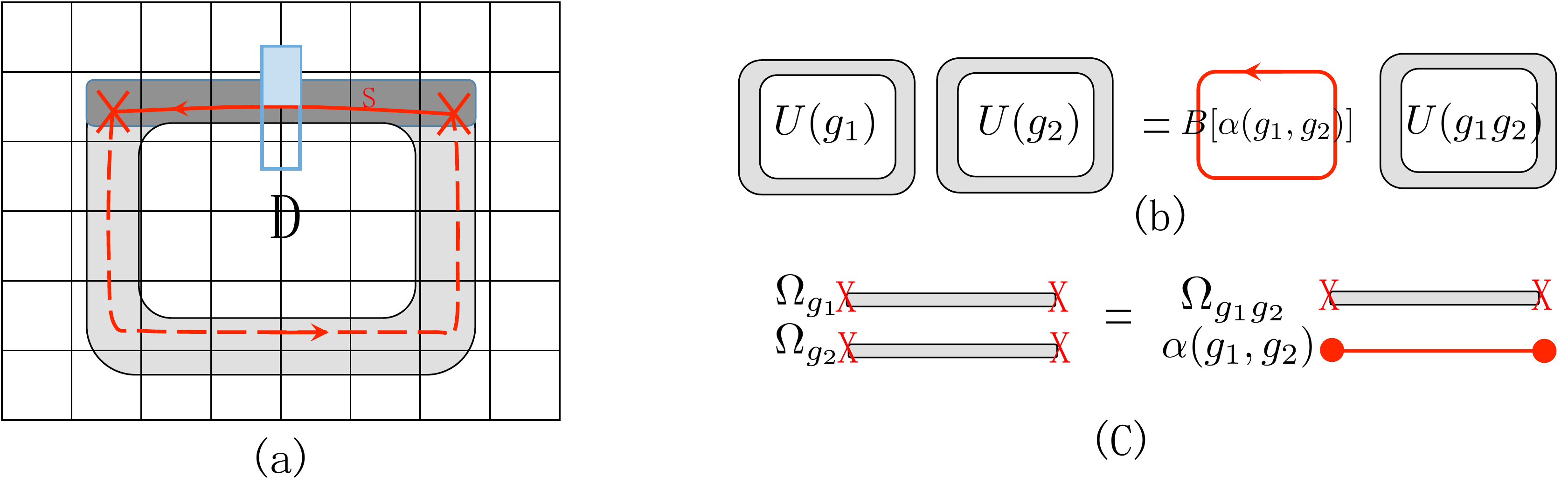}
\caption{(a) A pair of symmetry fluxes (red crosses) can be created at the end of a symmetry defect line $s$ (solid red line) by conjugating Hamiltonian terms across $s$ (e.g. the one in the blue box) with symmetry on one side of $s$. A full braiding of the symmetry fluxes around region $D$ is equivalent to applying symmetry locally to $D$ and the change in the ground state can be induced by applying unitary $U(g)$ along the boundary of $D$ (grey region). (b) $U(g)$ forms a projective representation of the symmetry. This is the same as Fig.\ref{proj} (b) except that we have made it clear that $U(g)$ acts only along the boundary of $D$. (c) The symmetry fluxes hence satisfy a projective fusion rule as given in Eq.\ref{proj_flux}.
}
\label{flux}
\end{center}
\end{figure}

Suppose that certain SF pattern can be realized in a lattice model as shown in Fig.\ref{flux}. We can insert a symmetry defect line $s$ of $g\in G$ (solid red line in Fig.\ref{flux}) by taking the Hamiltonian terms which are bisected by this line (for example the term in the blue box) and conjugate them by symmetry operator on the lattice sites to one side of the line (the shaded blue region).
\be
h_i \to \left(\prod_{k\in r.h.s.} M_k(g)\right)h_i \left(\prod_{k\in r.h.s.} M^{-1}_k(g)\right)
\ee
Here $r.h.s.$ stands for the right hand side of the defect line (or equivalently we can conjugate with $M_k(g^{-1})$ on the left hand side of the defect line). The terms which are not bisected by the defect line remain invariant. Near the end points -- the symmetry fluxes marked by red crosses in Fig.\ref{flux} -- there may be terms that are only partially bisected by the defect line, so it is ambiguous how to change them. But this ambiguity is not a problem, because we are only interested in how things change along the middle part of the defect line. As the Hamiltonian is changed along $s$, the ground state is also changed, but only within a finite width along $s$ (the dark grey region) due to the short range correlation in the state. Such a change in the ground state can be induced by applying a unitary operator $U^s(g)$ to the dark grey region along $s$ and we can think of this $U^s(g)$ as the string operator that creates a pair of symmetry fluxes and moves them around. An important difference between symmetry fluxes and anyons is that the symmetry fluxes are confined, i.e. the ground state changes along the full length of $s$ and costs an amount of energy (in terms of the original $H$) that is proportional to the length of $s$.


Now imagine creating a pair of symmetry fluxes, braiding them around region $D$ while changing the Hamiltonian along the way, and finally annihilating them (complete the circle along the dashed red line). After this full braiding process, all the Hamiltonian terms across the boundary of $D$ get conjugated by symmetry on one side while the terms inside and outside of $D$ remain invariant. The same change in Hamiltonian can be induced by applying symmetry locally to region $D$ ($\prod_{k\in D} M_k(g)$): the terms inside $D$ remain unchanged as they are invariant under global symmetry; the terms outside of $D$ remain unchanged as they are not acted upon; the terms across the boundary of $D$ get conjugated by symmetry on one side. Correspondingly, the change in the ground state can be induced by applying $\prod_{k\in D} M_k(g)$ to the region $D$. Therefore, braiding a pair of symmetry fluxes around a region corresponds to applying symmetry locally to that region.

On the other hand, the same change can be induced by applying a unitary string operator $U(g)$ along the boundary of $D$ (the grey region in Fig.\ref{flux}). Note that $U(g)$ is very different from $\prod_{k\in D} M_k(g)$ as it has no action deep inside $D$, although applying them to the ground state induces the same change. In fact, this $U(g)$ is exactly the local symmetry action we discussed in section \ref{cons} which satisfies the projective composition rule as given in Eq.\ref{proj_fs}, while the composition of $\prod_{k\in D} M_k(g)$ is not projective (because the composition of each $M_k(g)$ is not projective). Therefore, $U(g)$ carries the important information about the SF pattern while $\prod_{k\in D} M_k(g)$ does not. Eq.\ref{proj_fs} is saying that applying $U(g_1)$ and $U(g_2)$ is equivalent to applying $U(g_1g_2)$ up to the braiding of anyon $a(g_1,g_2)$ around the region $D$. This has to be the case because applying $U(g_1)U(g_2)U^{-1}(g_1g_2)$ to the Hamiltonian results in the same Hamiltonian, therefore the state can change at most by a phase factor. As $U(g_1)U(g_2)U^{-1}(g_1g_2)$ acts as a loop operator around region $D$, the phase factor can be induced by braiding an abelian anyon around $D$. On the other hand, from the fact that $U(g)$ implements the braiding of symmetry flux $\Omega_g$ around region $D$, we can further deduce that the fusion of $\Omega_{g_1}$ and $\Omega_{g_2}$ is equivalent to $\Omega(g_1g_2)$ up to $a(g_1,g_2)$, as shown in Eq. \ref{proj_flux} and Fig.\ref{flux} (c). Such a projective fusion rule among symmetry fluxes is the first step in several of the anomaly detection methods described below.

\subsection{Flux Fusion}

The flux fusion method discussed in \Ref{Hermele2015arxiv} can be used to detect anomalies in for example the $eCmT$ SF pattern\cite{,Vishwanath2013,Wang2013}. The $eCmT$ SF pattern exists in systems with a $Z_2$ topological order with anyons $\mathbb{1}$, $e$, $m$ and $f=e\times m$. Besides that, the system has charge conservation and time reversal symmetry $G=U(1) \times Z_2^T$. Note that here the charge conservation part is in direct product with the time reversal part, which means that the $U(1)$ charge gets reversed under time reversal. Therefore, it is better to think of the $U(1)$ part of the symmetry as spin conservation around the $z$ axis, which reverses direction under time reversal. 

The possibilities for symmetry fractionalization include fractional charges under $U(1)$ and Kramer doublet under time reversal ($\mathcal{T}^2=-1$). In the $eCmT$ case we consider the situation where $e$ carries half charge while $m$ carries integer charge, $e$ is a time reversal singlet while $m$ transforms as a Kramer doublet. The consistency condition of the SF pattern is satisfied implicitly with $f$ carrying half charge and transforming as a Kramer doublet.


In the first step of the anomaly test, we introduce symmetry flux of the $U(1)$ part of the symmetry and study its projective fusion rule (hence the name flux fusion). As $e$ has half charge while $m$ has integer charge, we can conclude that
\be
\Omega_{\pi} \times \Omega_{\pi} = m \ \ \text{or} \ \ a(\pi,\pi) = m
\ee

In the second step, we ask the question: how does $\Omega_{\pi}$ transform under time reversal symmetry? First of all, we observe that time reversal does not change the amount of flux, because time reversal commutes with $U(1)$ rotations. Therefore, $\Omega_{\pi}$ transforms either as a time reversal singlet or a Kramers doublet. But this is in contradiction to $m$ being a Kramers doublet because whether $\Omega_{\pi}$ is a time reversal singlet or doublet, the composite $\Omega_{\pi}\times \Omega_{\pi}=m$ must be a singlet. We have thus found that $eCmT$ is an anomalous fractionalization pattern.

The flux fusion method hence consists of 1. find the projective fusion rule of symmetry fluxes 2. examine whether symmetry fractionalization on the symmetry fluxes can be consistent with the fusion rule. It applies to a variety of situations including those with time reversal and spatial symmetries, but it is also restricted and does not apply when, for example, the symmetry flux type is changed under symmetry transformation.

In \Ref{Metlitski2013, Wang2013, Wang2014, Wang2016}, anomalies in SF patterns with $U(1)$ symmetry are detected using a `monopole tunneling' method. Imagine tunneling a monopole through the 2D system and leaving behind a $2\pi$ flux. If the $2\pi$ flux carries projective symmetry number or statistics, then the quantum number of the monopole will change in a nontrivial way after tunneling, hence indicating anomaly. \Ref{Metlitski2013} first used this method to show the anomaly of the $eCmC$ state where the $2\pi$ flux is a fermion (unless there is an odd Hall conductance). In \Ref{Wang2013}, it was shown that for $eCmT$ the $2\pi$ flux is a Kramer doublet and hence nontrivial. The flux fusion method is equivalent to this method when the symmetry flux inserted is a $U(1)$ flux.


\subsection{Conflicting Symmetries}

In \Ref{Cho2014} and \Ref{Kapustin2014}, it was proposed that anomalies in SF patterns with $G_1\times G_2$ type of symmetry may be detected through the `conflict' between $G_1$ and $G_2$ which is revealed through the breaking of $G_2$ by gauging $G_1$. One example discussed in \Ref{Cho2014, Bi2015} has $Z_2$ topological order and $Z^A_2\times Z^B_2$ unitary symmetry with group elements $\{I,g_A,g_B,g_Ag_B\}$. The anyons transform as
\be
e: U_e(g_A) = i\sigma_z, U_e(g_B)=\sigma_x; \ \ m: U_m(g_A) = I, U_m(g_B) = i\sigma_y
\ee
The $f$ anyon transforms as $e$ and $m$ combined, so the  SF pattern is consistent. The fact that both $e$ and $m$ transform nontrivially under the symmetry lead to potential anomaly\cite{Kapustin2014arxiv}.

If we look at the two $Z_2$ symmetry subgroups individually, their actions are simple and not anomalous. For example, under the $Z^A_2$ subgroup (generated by $g_A$), $m$ transforms trivially while $e$ carries half charge ($\left(U_e(g_A)\right)^2=-I$). Such a symmetry action can be realized in a strictly 2D system. Correspondingly we can fully gauge the $Z^A_2$ subgroup and obtain a larger topological order. After gauging, the $Z^B_2$ part remains a global symmetry of the system and acts on the anyons in the larger topological order. The `conflict' between $Z^A_2$ and $Z^B_2$, and hence the anomaly of the SF pattern, is detected through the observation that $Z^B_2$ does not act on the larger topological order in a consistent way.

In particular, gauging $Z_2^A$ results in a $Z_4$ topological order with elementary gauge charge $e_4$ ($e_4^4=\mathbb{1}$) and elementary gauge flux $m_4$ ($m_4^4=\mathbb{1}$). $e_4$ and $m_4$ has a mutual statistics of $i$. $e_4$ comes from the $e$ anyon of the original $Z_2$ topological order; $e_4^2=c$ is the symmetry charge of the $Z_2^A$ symmetry; $m_4$ comes from the symmetry flux $\Omega_A$ of $Z_2^A$; $m_4^2=m$ is the $m$ anyons of the original $Z_2$ topological order. 

In the next step, one can ask how the $Z_2^B$ symmetry acts on the $Z_4$ topological order. Because in the original SF pattern, $U_e(g_B) = \sigma_x$ flips between the two components of $e$ which differ by a $-1$ under the action of $U_e(g_A)$ (one with eigenvalue $i$ and one with $-i$), in the gauged theory $g_B$ exchanges $e_4$ and $e_4c=e_4^3$. In order to keep the statistics of the $Z_4$ topological order invariant, $g_B$ has to exchange $m_4$ and $m_4^3$. However, this is not consistent with the fact that $m=m_4^2$ carries half charge under $g_B$. In order to see this, we need the consistency condition of SF patterns when anyon types are changed under the symmetry. We did not discuss this in this review, although the result in section\ref{cons} can be generalized directly as shown in \Ref{Barkeshli2014arxiv,Tarantino2015arxiv}. In \Ref{Cho2014}, the SF pattern and the gauging procedure was discussed using the Chern-Simons formalism and it was explicitly shown that the $Z_2^B$ symmetry action is not consistent with the $Z_4$ topological order obtained after gauging $Z_2^A$.

\subsection{Gauging Obstruction}
\label{gOb}

A powerful mathematical method exists to detect anomalies with unitary on-site symmetries. This is discussed in \Ref{Etingof2010} as the `obstruction to the extension of a braided fusion category by a finite group'. In \Ref{Chen2014}, the projective semion example was used to illustrate this method. 

The projective semion example has the topological order of a chiral semion theory (reviewed in Appendix A), whose only nontrivial anyon $s$ has topological spin $i$. The system also has unitary $Z_2\times Z_2$ symmetry with group elements $\{I,g_x,g_y,g_z\}$. Four possible SF patterns on the semion are:
\be
\begin{array}{llll}
PS_0: & U_s(g_x) = i\sigma_x, & U_s(g_y) = i\sigma_y, & U_s(g_z) = i\sigma_z \\
PS_X: &  U_s(g_x) = i\sigma_x, & U_s(g_y) = \sigma_y, & U_s(g_z) = \sigma_z \\
PS_Y: &  U_s(g_x) = \sigma_x, & U_s(g_y) = i\sigma_y, & U_s(g_z) = \sigma_z \\
PS_Z: &  U_s(g_x) = \sigma_x, & U_s(g_y) = \sigma_y, & U_s(g_z) = i\sigma_z
\end{array}
\label{ps}
\ee
Correspondingly, the symmetry fluxes fuse projectively as
\be
\begin{array}{llllll}
PS_0: &  a(X,X)=s, & a(Y,Y)=s, & a(Z,Z)=s, & a(X,Y) = a(Y,Z) = a(Z,X) = s, & a(Y,X) = a(Z,Y) = a(X,Z) =\mathbb{1}\\
PS_X: &  a(X,X)=s, & a(Y,Y)=\mathbb{1}, & a(Z,Z)=\mathbb{1}, & a(X,Y) = a(Y,Z) = a(Z,X) = s, & a(Y,X) = a(Z,Y) = a(X,Z) =\mathbb{1}\\
PS_Y: &  a(X,X)=\mathbb{1}, & a(Y,Y)=s, & a(Z,Z)=\mathbb{1}, & a(X,Y) = a(Y,Z) = a(Z,X) = s, & a(Y,X) = a(Z,Y) = a(X,Z) =\mathbb{1}\\
PS_Z: &  a(X,X)=\mathbb{1}, & a(Y,Y)=\mathbb{1}, & a(Z,Z)=s, & a(X,Y) = a(Y,Z) = a(Z,X) = s, & a(Y,X) = a(Z,Y) = a(X,Z) =\mathbb{1}\\
\end{array}
\label{pspf}
\ee

Now, if the symmetry in the projective semion theory can be consistently gauged, we should be able to define for the fluxes not only the projective fusion rules but also the braiding and fusion statistics involved with exchanging two fluxes or fusing three of them in different orders. These statistics cannot be chosen arbitrarily, but have to satisfy certain consistency conditions\cite{Barkeshli2014arxiv}. Failure to satisfy these consistency conditions reveals the anomaly in the SF pattern. We are not going to explain the reasoning which led to the conclusion in \Ref{Etingof2010} but only to quote that, to determine whether these consistency conditions can be satisfied, it suffices to calculate the following quantity
\be
\begin{array}{ll}
\nu(f,g,h,k) = &R_{a(h,k),a(f,g)} F_{a(g,h),a(f,gh),a(fgh,k)}F^{-1}_{a(g,h),a(gh,k),a(f,ghk)}F_{a(f,g),a(h,k),a(fg,hk)}\\
&F^{-1}_{a(f,g),a(fg,h),a(fgh,k)} F_{a(h,k),a(g,hk),a(f,ghk)}F^{-1}_{a(h,k),a(f,g),a(fg,hk)}
\end{array}
\label{ob}
\ee
The $F$ and $R$ symbols depend on the anyon coefficient of the projective fusion rule of the symmetry fluxes and their value are determined from the semion topological order (as reviewed in Appendix A). $\nu(f,g,h,k)$ is a phase factor that depends on four group elements $f,g,h,k \in G$. However, in some cases it is possible that $\nu(f,g,h,k)$ can actually be generated from a phase factor $\mu$ which depends only on three group elements as
\be
\nu(f,g,h,k) = \mu(g,h,k) \mu^{-1}(fg,h,k)\mu(f,gh,k)\mu^{-1}(f,g,hk)\mu(f,g,h)
\label{cbound}
\ee
The powerful conclusion of \Ref{Etingof2010} is that: if this is the case, the SF pattern is not anomalous and the gauging process can go through. However, if this is not true, that is if $\nu(f,g,h,k)$ cannot be decomposed as in Eq.\ref{cbound}, then the SF pattern is anomalous and there is obstruction to the gauging procedure. If we do this calculation for the projective semion SF pattern listed above, we can find that $PS_0$ is non-anomalous and $PS_X$, $PS_Y$, $PS_Z$ are anomalous.

This `Gauging Obstruction' method hence provides a generic tool for detecting anomalies in SF patterns with unitary symmetries. It is possible that this method can be generalized to anti-unitary symmetries\cite{Chen2014}, although the idea of `gauging' applies most naturally to unitary symmetries. 

\section{Non-anomalous SF pattern as 2D spin liquid}
\label{spliquid}

Combining the consistency condition and the anomaly detection methods, we can now try to enumerate all SF patterns that can be realized in strictly 2D models with a given topological order and given global symmetry. In section \ref{2Dexample}, we review one such effort for Kagome lattice chiral spin liquid -- a system with chiral semion topological order and symmetries common to a spin model on Kagome lattice (spin rotation, lattice symmetry, etc). Which one of these possible SF patterns is actually realized in the physically realistic models, like the nearest neighbor Heisenberg model? To answer this question, we need to be able to detect SF patterns through numerical or experimental probes. Some of the proposed probing methods are reviewed in section \ref{2Dnum} and section \ref{2Dexp} respectively.

\subsection{Classification Example}
\label{2Dexample}

\Ref{Cincio2015arxiv} and \Ref{Zaletel2016} considered the classification of chiral spin liquid on the Kagome lattice. The chiral spin liquid contains one nontrivial anyon -- the semion $s$, which can carry fractional representations of $SO(3)$ spin rotation, translation and plaquette centered inversion symmetry. While in a chiral spin liquid time reversal and reflection symmetries are individually broken, their combined action can still be preserved, under which the semion can transform projectively. It was found in \Ref{Cincio2015arxiv} and \Ref{Zaletel2016} that in the physically interesting case of an odd number of spin $1/2$s per unit cell, there is only one non-anomalous SF pattern for the chiral spin liquid.

First, it was proven that in a lattice with an odd number of spin $1/2$s per unit cell, at least one anyon has to carry half integer spin\cite{Cheng2015arxiv}. In this case, it has to be the semion. Here we want to comment that half integer spin is a nontrivial fractionalization pattern for the anyons even though the system is composed of half integer spins per lattice site. This is because, as long as the Hilbert space on each lattice site is not a direct sum of half integer and integer spins, global excitations (the difference between excited state and ground state) always carry integer spin. Only fractional excitations like the anyons can carry half integer spins.

In the next step, fractional quantum numbers are determined for the semion under translation symmetry $T_x$, $T_y$, inversion symmetry $I_v$ and the combined action of reflection and time reversal $R_x$, $R_y$. It was found that the only non-anomalous possibility is that
\be
U(I_v)^2 = s, U(T_x)U(T_y)U(T_x^{-1})U(T_y^{-1})=s, U(R_x)^2 = s, U(R_y)^2=s
\ee
which is saying that, for example, applying inversion ($I_v$) twice on the semion results in the phase factor of $-1$(the braiding statistics between two semions). Therefore, the SF pattern of the chiral spin liquid on Kagome lattice is completely fixed by the consistency condition and the anomaly free condition.

\subsection{Numerical Probe}
\label{2Dnum}


\begin{wrapfigure}{r}{0.5\textwidth}
  \vspace{-20pt}
  \begin{center}
    \includegraphics[width=0.25\textwidth]{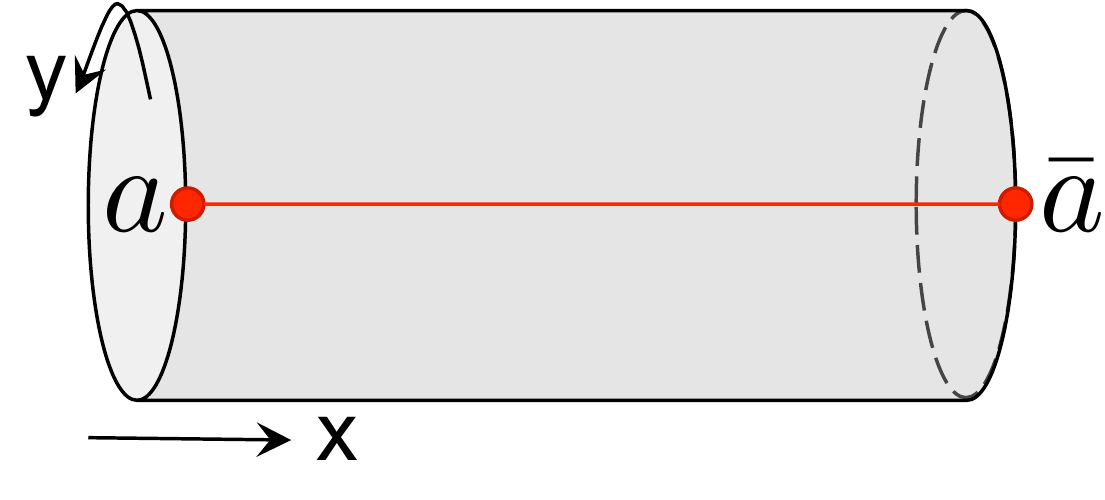}
  \end{center}
  \vspace{-20pt}
  \caption{Putting the 2D system on a cylinder with finite circumference effectively reduces the system to 1D; creating a pair of anyons $a$ and $\bar{a}$ at the two ends of the cylinder can change the symmetry protected topological order of the 1D state which can be measured to determine the SF pattern of $a$.}
  \label{cylinder}
  \vspace{-10pt}
\end{wrapfigure}

In other types of spin liquids (with different topological order or lattice symmetry), usually more than one SF pattern is potentially realizable\cite{Lu2011,Lu2011a,Messio2013,Zheng2015arxiv,Qi201509arxiv,Bieri2016,Lu2016, Qi2016arxiv}. To determine which one is actually realized in a specific model, one can perform tests in numerical simulations, as discussed in \Ref{Huang2014,Cincio2015arxiv,Zaletel2015arxiv,WangL2015,Qi2015,Zaletel2016,Saadatmand2016arxiv}. 

A useful geometry for this purpose is the cylinder, as shown in Fig.\ref{cylinder}, with periodic boundary condition in the $y$ direction and open boundary condition in the $x$ direction. Consider the situation where the circumference along the $y$ direction is finite while the length along the $x$ direction is infinite. The 2D topological state with symmetry is then effectively reduced to a 1D gapped state with symmetry, with 1D symmetry protected topological order. This process is called `dimensional reduction'. In Appendix B, we give a brief review of the notion of symmetry protected topological order. 

Now to detect symmetry fractionalization, one can create a pair of anyons $a$ and $\bar{a}$ and bring them to the two ends of the cylinder. This process can change the symmetry protected topological order of the dimensional reduced system and the SF pattern of $a$ is encoded in the change of the edge state of the 1D system. By measuring the symmetry protected topological order of the dimensional reduced system with, e.g., nonlocal order parameters\cite{Pollmann2012,Haegeman2012}, one can determine the SF pattern of different anyons.

\subsection{Experimental Probe}
\label{2Dexp}

In fqH systems, the fractional charge carried by the anyons has been directly measured through shot-noise and local tunneling experiements\cite{Wen1990, de-Picciotto1997, Saminadayar1997, Martin2004}. For spin liquids, candidate systems have been identified in organic Mott insulator $\kappa$-(BEDT-TTF)$_2$Cu$_2$(CN)$_3$\cite{Shimizu2003} and herbertsmithite ZnCu$_3$(OH)$_6$Cl$_2$\cite{Helton2007} etc. The nonexistence of magnetic order in these materials at very low temperature gives strong evidence that they are in a spin liquid state\cite{Hastings2004}. Can we obtain more direct evidence for the existence of a spin liquid by, for example, identifying the SF pattern? Several proposals have been made to address this issue.

\Ref{Senthil2001} discussed different ways to detect spin charge separation in a spin liquid. In particular,  in a superconductor -- insulator -- superconductor junction, if the insulator contains chargons -- spinless charge $e^{-}$ bosons, the ac Josephson current obtained by applying a dc voltage $V$ across the junction will have an oscillating component at frequency $\omega = e^{-}V/\hbar$ in addition to the component at $\omega=2e^{-}V/\hbar$. Similarly, the tunneling conductance into small superconducting islands across an insulating barrier behaves differently if the insulating barrier has spin-charge separation. With non-fractionalized insulating barrier, the tunneling conductance vary with the total charge on the island with a period of $2e^{-}$. If the the barrier contains chargons, the period becomes $e^{-}$. 

More recently, \Ref{Barkeshli2014} proposed that spin-charge separation can be directly measured by tunneling electrons into the spin liquid through suitably chosen boundary state. In particular, if chargons (spinons) are condensed at the boundary, electrons can leave their charge (spin) behind and enter the spin liquid as a fractional particle. This can be detected through the oscillation of the local density of states of the tunneling electron with the applied voltage, which results from the coherent propagation of fractional particles across the spin liquid.

On the other hand, \Ref{Essin2014} discussed experimental signature of crystal momentum fractionalization in $Z_2$ spin liquid. For example, translation in the $x$ and $y$ direction can anti-commute on a spinon
\be
U_e(T_x)U_e(T_y)U_e(T_x^{-1})U_e(T_y^{-1}) = -1
\ee
That is, bringing the spinon around a plaquette in the square lattice results in a phase factor of $-1$. If this is the case, then \Ref{Essin2014} showed that the density of states of the two-spinon continuum in the spectrum has an enhanced periodicity. In particular, for any two spinon state $|a\rangle$, applying translation on only one of the spinon generates three other different states $U_{e_1}(T_x)|a\rangle$, $U_{e_1}(T_y)|a\rangle$, $U_{e_1}(T_x)U_{e_1}(T_y)|a\rangle$ with different lattice momentum but the same energy. Therefore, the density of states repeats itself four times in the Brillouin zone, which can be detected through neutron scattering experiments.

\section{Anomalous SF pattern on surface of 3D systems}
\label{surface}

\begin{wrapfigure}{r}{0.5\textwidth}
  \vspace{-20pt}
  \begin{center}
    \includegraphics[width=0.35\textwidth]{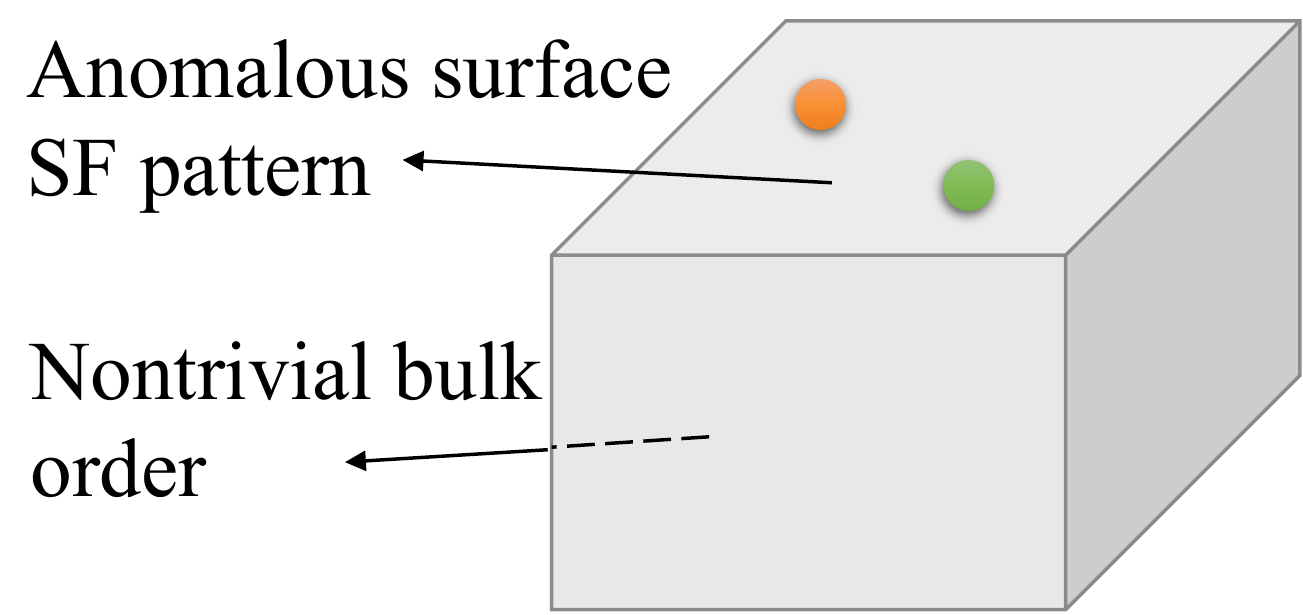}
  \end{center}
  \vspace{-20pt}
  \caption{Anomalous SF patterns can be realized on the surface of a 3D system and their anomaly is a reflection of the nontrivial order in the 3D bulk.}
  \label{3D}
  \vspace{-10pt}
\end{wrapfigure}

When the SF patterns are anomalous, they cannot be realized in strictly 2D systems. However, they are not completely impossible either. It was realized that they can be found on the surface of a 3D system and their anomaly is a reflection of the nontrivial order inside the 3D bulk, as shown in Fig.\ref{3D}. This connection was first pointed out in \Ref{Vishwanath2013} and many examples have been worked out demonstrating this kind of bulk-boundary correspondence. We review several interesting cases in this section, with anomalous SF pattern on the surface and symmetry protected topological order in the bulk. For a brief review on symmetry protected topological order, see Appendix B.

\subsection{Bosonic Topological Insulator}

First, it was realized\cite{Vishwanath2013,Wang2013} that the $eCmC$ SF pattern with time reversal discussed in section \ref{anom:example} can be realized on the surface of a 3D bosonic topological insulator. Being on the surface of a 3D system, it avoids the contradiction discussed in section \ref{anom:example} where the nonzero Hall conductance on the edge explicitly breaks time reversal. This is due to a simple, yet powerful, geometric observation that the 2D boundary of a 3D bulk does not have a boundary of its own. Therefore, we will not be able to observe the nonzero Hall conductance and there is no contradiction to the system being time reversal invariant. On the other hand, its existence as the 2D surface indicates special order in the 3D bulk, even though the anyons only live on the surface and cannot tunnel into the bulk. In particular it was shown\cite{Metlitski2013} that the monopole in the 3D bulk is a fermion -- the `statistical Witten effect', indicating that the bulk state cannot be smoothly connected to a product state without breaking charge conservation and time reversal symmetry. The bulk state is the so-called `Bosonic topological insulator' , following the terminology of fermionic topological insulators\cite{Fu2007,Moore2007,Roy2009}.

\subsection{Bosonic $Z_2\times Z_2$ symmetry protected topological phases}

The projective semion SF patterns discussed in section \ref{gOb} are realized on the surface of 3D bosonic symmetry protected topological phases with $Z_2\times Z_2$ symmetry. It was shown in \Ref{Chen2014} that the $PS_X$, $PS_Y$ and $PS_Z$ SF patterns are realized on the surface of three different nontrivial phases. A strong indication of the relation between the surface anomaly and the bulk symmetry protected topological order is that they are both characterized by $\nu(f,g,h,k)$ (Eq.\ref{ob}), the fourth cocycle of the group. This kind of bulk-boundary correspondence is expected to apply for all unitary groups\cite{WangCJ2016}. Moreover, one of the 3D bosonic symmetry protected topological phases with time reversal symmetry seems to fit into this scheme as well whose surface can have $Z_2$ topological order with both $e$ and $m$ transforming under time reversal as a Kramer doublet\cite{Burnell2014}. If we use Eq.\ref{ob} to compute $\nu(f,g,h,k)$, we would obtain the nontrivial cocycle related to the bulk order.

\subsection{Fermionic Topological Insulator}

\begin{figure}[htbp]
\begin{center}
\includegraphics[height=4.0cm]{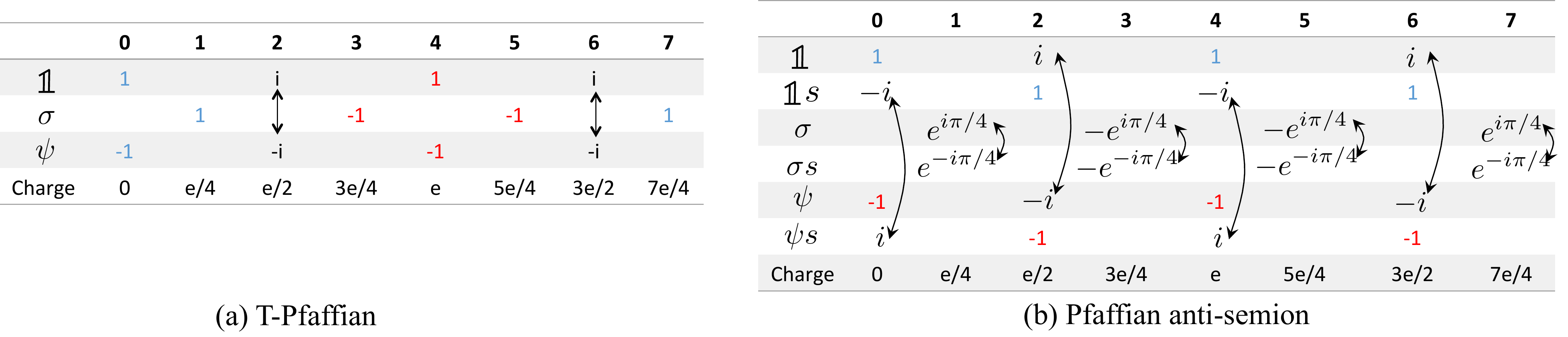}
\caption{The SF pattern on the surface of a 3D fermionic topological insulator. (a) the T-Pfaffian topological order is composed of an Ising sector ($\mathbb{1}$, $\sigma$, $\psi$) and a $U(1)_8$ sector ($0\sim 7$) (b) the Pfaffian anti-semion topological order is composed of an Ising sector, an anti-semion sector ($s$) and a $U(1)_8$ sector. Entries in the table are topological spins of the anyon. Anyon pairs connected with an arrow are time reversal partners. Blue entries are for time reversal singlets while red entries are for time reversal doublets. The charge carried by the anyons are given in the last row of the tables.}
\label{TI_SF}
\end{center}
\end{figure}

It is of great interest to investigate what kind of SF pattern can exist on the surface of 3D fermionic topological insulator, whose bulk state has been realized experimentally\cite{Hsieh2008,Hsieh2009,Chen2009}. \Ref{Bonderson2013, Wang2013a, Chen2014b, Metlitski2015} addressed this question and found the following two answers: the T-Pfaffian state and the Pfaffian-anti-semion state.  Table \ref{TI_SF} summarizes the anyon content and the SF pattern of the two states.

The two states share the following common properties: 1. as surface states, they have both time reversal and charge conservation symmetry; 2. if realized in 2D with charge conservation symmetry, the system would have Hall conductance $\sigma_{xy}=\frac{\left(e^-\right)^2}{2h}$ and hence break time reversal symmetry; 3. if superconductivity is induced in the states, the $\pi$ flux would host a Majorana zero mode. All these ensures that they are consistent with the bulk order. Although it suffices to have non-interacting electrons to realize the bulk state, to induce topological order and symmetry fractionalization on the surface, strong interaction is necessary. Exactly what kind of interaction is needed is still under investigation. By considering possible surface SF patterns, a complete classification of interacting topological insulators has been obtained\cite{Wang2014}. Therefore, the study of SF pattern on the surface provides a useful strongly interacting perspective of symmetry protected topological phases. More discussions of these surface states can be found in \Ref{Metlitski2015arxiv, Seiberg2016arxiv, Mross2015}

\subsection{Fermionic Topological Superconductor}

\Ref{Fidkowski2013,Metlitski2014arxiv,Metlitski2015arxiv,Witten2016arxiv} studied the SF pattern possible on the surface of fermionic topological superconductors and the findings are listed in table \ref{TSc_SF}. An interesting observation obtained from this analysis is that, when $\nu=16$ (i.e. 16 copies of the elementary topological superconductor), the SF pattern becomes non-anomalous, implying that the $\nu=16$ topological superconductor can be smoothly connected to a trivial $s$-wave superconductor. Such a connection is of course only possible with strong interaction, as in the non-interacting classification each integer $\nu$ corresponds to a different topological superconductor\cite{Schnyder2008,Kitaev2009}. A similar conclusion is reached for topological crystalline insulators where the $\nu=8$ state is found to be trivial through the study of surface SF pattern\cite{Qi2015a}.

\begin{table}[h]
\be
\begin{array}{|c|c|c|c|c|c|}
\hline
   \text{TSc}  & \text{Anyon content} & \text{Fusion rule} & \text{Topological spin} & \mathcal{T} \text{action on anyons} & \mathcal{T}^2 \text{action on anyons} \\ \hline
\nu=1 \text{(mod $2$)}  & \mathbb{1}, S,Sf,f & S\times S = \mathbb{1}+S+Sf  & \theta_S = i & S \leftrightarrow Sf & \\ \hline
\nu=\pm 2  & \{\mathbb{1},s\}\{\mathbb{1},f\} & s\times s = \mathbb{1} & \theta_s = i & s \leftrightarrow sf & \mathcal{T}^2 = \pm i \text{\ on\ } s \\ \hline
\nu=4 &  \{\mathbb{1},s_1\}\{\mathbb{1},s_2\}\{\mathbb{1},f\} & s_i\times s_i = \mathbb{1}, i =1,2 & \theta_{s_{1,2}} = i &s_{1,2} \leftrightarrow s_{1,2}f & \mathcal{T}^2 = \pm i \text{\ on \ } s_1,\ s_2  \\ \hline 
\nu=8  & \{\mathbb{1}, F_1,F_2,F_3\}\{\mathbb{1},f\} & F_i\times F_i = \mathbb{1}, F_1\times F_2 = F_3 & \theta_{F_i} = -1 & F_i \leftrightarrow F_i & \mathcal{T}^2=1 \text{\ on \ } F_i\\ \hline 
\end{array}
\ee
\caption{The SF pattern on the surface of 3D fermionic topological superconductors. $f$ is the Bogoliubov quasi-particle of the superconductor. $\mathcal{T}$ stands for time reversal.}
\label{TSc_SF}
\end{table}

\section{Conclusion and Outlook}
\label{discussion}

In this paper, we reviewed recent progress in answering the following question: given a set of anyons and certain global symmetry, what are the possible fractional ways the anyons can transform under the symmetry? The answer comes in two steps. First, for a symmetry fractionalization pattern to be possible at all, it needs to be consistent with the anyon fusion rule as discussed in section\ref{cons}. Secondly, among all the possible SF patterns, some are non-anomalous and can be realized in strictly 2D models (as discussed in section\ref{spliquid}) while others are anomalous and can be realized only on the surface of a 3D system (as discussed in section \ref{surface}). Methods which can be used to distinguish the anomalous ones from the non-anomalous ones are discussed in section \ref{anomaly}.

There are many aspects of this topic which we would like to but are not able to cover in this review: 1. we discussed the possibilities to realize the SF patterns either in 2D or on 3D surface, but we did not present any concrete models. A parton construction for the spin liquids can be obtained following \Ref{Wen2002,Wang2006}. More lattice models can be found in \Ref{Levin2011,Hermele2014,Hao2015,Lee2016arxiv} and some field theory constructions are discussed in \Ref{Levin2009,Lu2016SET,Hung2013,Hung2013_Kmatrix,Lu2014}. 2. We focused mostly on cases where anyons types are not changed by the symmetry. The consistency condition and some of the anomaly test can also be generalized to the cases where anyon types are changed\cite{Barkeshli2014arxiv,Tarantino2015arxiv,Cui2016}. We want to mention one particularly useful consistency condition when time reversal exchanges anyon types\cite{Wang2013a,Bonderson2013}: if $a$ and $\mathcal{T}(a)$ are exchanged under time reversal symmetry, then the time reversal action on $b$ -- their fusion product -- depends on the braiding statistics of $a$ and $\mathcal{T}(a)$; in particular, $U_b(\mathcal{T})U_b^*(\mathcal{T})=\theta^b_{a\mathcal{T}(a)}=\theta_b$\footnote{This is only true when $N^b_{a\mathcal{T}(a)}$ is odd -- private communication with Parsa Bonderson}. 3. In section\ref{surface}, we discussed the realization of anomalous SF pattern on the surface of symmetry protected topological phases. It is also possible to have anomalous SF pattern on the surface of other types of topological phases. \Ref{Fidkowski2015arxiv} studied the SF pattern (with $D_{16}$ gauge theory and $Z_2$ global symmetry) on the surface of a 3D symmetry enriched topological phase with a $H^3(G,A)$ type anomaly as discussed in \Ref{Etingof2010,Barkeshli2014arxiv}. An example with $Z_3$ gauge theory and $Z_3\times Z_3$ symmetry was studied in \Ref{Kapustin2014,Kapustin2014arxiv} which was found to have a $H^3(G,Z_3)$ type anomaly related to the non-surjective mapping between Dijkgraaf-Witten gauge theories\cite{Dijkgraaf1990}. It was shown to be realizable on the surface of 3D symmetry protected topological phases with 2-group symmetry\cite{Thorngren2015arxiv}. This is among a class of abelian Chern-Simon theories with 't Hooft anomaly studied in \Ref{Kapustin2014,Kapustin2014arxiv}, with different anomaly generating mechanisms at play (nontrivial fractionalization on both gauge charge and gauge flux, non-surjective mapping between Dijkgraaf-Witten gauge theories, a combination of the two, etc).

With all the theoretical progress, we are one step closer to seeing symmetry fractionalization in experiments other than the fractional quantum Hall systems. However, many problems remain to be addressed. For example, which SF pattern is actually realized with the physically realistic Heisenberg model, what kind of interaction is needed to realize the anomalous SF patterns on the surface of Topological insulator and superconductors, what exactly can be measured to identify the SF pattern, etc. On the other hand, SF patterns in 3D topological phases show new features compared to 2D but our understanding of them is far from complete. Fractional excitations in 3D appear as monopoles or loops. How symmetry fractionalizes on them is an interesting open question\cite{Cheng2015arxiv3D,Chen2016arxiv}. Finally, symmetry action in gapless systems can be very different from that in gapped systems. It is interesting to understand how symmetry fractionalizes in correlated gapless systems\cite{Bonderson2016arxiv}.

\begin{acknowledgments}
XC is grateful to Parsa Bonderson, Meng Cheng, Lukasz Fidkowski, Mike Hermele, Anton Kapustin, Yuan-Ming Lu, Ryan Thorngren, Senthil Todadri, Ashvin Vishwanath, Chong Wang and Mike Zaletel for carefully reading the paper and providing valuable feedback. XC is supported by the Caltech Institute for Quantum Information and Matter and the Walter Burke Institute for Theoretical Physics.
\end{acknowledgments}


\appendix
\section*{Appendix A: Two dimensional anyon theory}

In this section, we give a very brief review on two dimensional anyon theory. For a more extensive introduction to the topic, see \Ref{Kitaev2006}.

In two dimensional gapped topological phases, anyons are point excitations that cannot be removed by acting locally around the point. For example, the elementary excitations with $1/3$ electron charge in the $\nu=1/3$ fractional quantum Hall (fqH) is an anyon. Two point excitations that can be mapped into each other through local operation are considered to be the same type of anyon. In the case of fqH, as individual electrons can be added or removed from the system locally, point excitations which are $1/3 + n$ ($n\in \mathbb{Z}$) fractions of an electron belong to the same type of anyon. Point excitations that can be created locally are equivalent to no excitation and are called the trivial anyon (denoted as $\mathbb{1}$). All the nontrivial anyons must be created in particle-antiparticle pairs, but then can be separated far apart from each other without extensive energy cost.  

Suppose that a topological phase contains anyon types $a$, $b$, $c$, etc. The anyon types obey a fusion rule given in general by
\be
a \times b = \sum_c N^c_{ab} c
\ee
where $N^c_{ab}$ are nonnegative integers. For every $a$, $a\times \mathbb{1}=a$. Moreover, for each $a$, there exists a unique $\bar{a}$ such that $N^{\mathbb{1}}_{a\bar{a}}\neq 0$. 

Each anyon type $a$ is characterized by a positive number $d_a$ called the quantum dimension. The physical meaning of $d_a$ is such that if the system contains $M$ type $a$ excitations which are separated far away from each other, then the degeneracy of the system scales as $d_a^M$ as $M$ goes to infinity. The quantum dimension of different anyon types satisfy
\be
d_a \cdot d_b = \sum_c N^{c}_{ab} d_c
\ee
When $d_a=1$, $a$ is said to be abelian. When $d_a>1$, $a$ is nonabelian. 

The most important property of anyons is their fractional exchange and braiding statistics. Suppose that we take two anyons of the same type $a$ and exchange their position. As these two anyons are identical particles, exchanging them may change the global wave function only by a phase factor (if $a$ is nonabelian, we need to fix their fusion product and we only consider cases without fusion multiplicity). If we denote this phase factor as $\theta^{c}_a$, then
\be
\theta_a = d_a^{-1} \sum_c d_c \theta^{c}_a
\ee 
is also a phase factor (which is a nontrivial fact) and is called the topological spin of $a$. Anyons with $\theta_a = 1, -1, i$ are called bosons, fermions, semions respectively. The possibility of $\theta_a$ being different from $1$ or $-1$ led to the terminology of `any-on'\cite{Wilczek1982}.

If we take two (possibly different) types of anyon $a$ and $b$ and bring one around another in full circle (with fixed fusion product $c$ for nonabelian $a$, $b$), this process gives rise to the braiding statistics between $a$ and $b$ and is related to the topological spin as
\be
\theta^c_{ab} = \frac{\theta_c}{\theta_a\theta_b}
\ee 

If $a$ is abelian, then
\be
\theta^{c_1}_{ab_1} \theta^{c_2}_{ab_2} = \theta^{c}_{ab}
\ee
where $N^b_{b_1b_2} > 0$ and $c_1$, $c_2$, $c$ are uniquely fixed by $b_1$, $b_2$, $b$ because $a$ is abelian. This is just saying that the braiding statistics of $b_1$, $b_2$ around $a$ is additive when $a$ is abelian.

The data set regarding possible anyon types, their fusion rules, and their exchange and braiding statistics specifies a long range entangled topological order. In two dimensional topological phases, each nontrivial anyon type has nontrivial braiding statistics ($\neq 1$) with at least one type of anyon. Because of this, the anyons are called `fractional' excitations. The fractionalization of braiding and exchange statistics exists independent of possible global symmetries in the system.

The full algebra of anyon fusion and braiding can be described with two sets of data: the $\left[F^{abc}_d\right]_{e,f}$ symbol which depends on six anyon labels and the $R^{ab}_c$ symbol which depends on three anyon labels. $\left[F^{abc}_d\right]_{e,f}$ is related to the non-associativity of the fusion rule: fusing $a,b$ into $e$ first and then fusing $e,c$ into $d$ differs from fusing $b,c$ into $f$ first and then fusing $a,f$ into $d$ by $\left[F^{abc}_d\right]_{e,f}$. $R^{ab}_c$ is related to the half braiding process between $a$ and $b$ when their fusion product is $c$. To describe a valid topological theory, they have to satisfy certain consistency conditions -- the pentagon and hexagon equations. For details regarding these consistency conditions, see \Ref{Kitaev2006}. Unlike topological spin and braiding statistics, $F$ and $R$ are not invariants of the topological phase and may change under certain basis transformations. They are related to the invariant quantities as
\be
R^{aa}_c = \theta^c_a, \ \ R^{ab}_cR^{ba}_c = \theta^c_{ab}
\ee
If the anyons are all abelian, then $F$ depends on only $a,b,c$ and $R$ depends on only $a,b$.

Some simple examples:

A chiral semion topological order contains only one type of nontrivial anyon: the semion $s$. $s$ is abelian and
\be
s\times s = \mathbb{1}
\ee
where $\mathbb{1}$ denotes the vacuum sector of the topological phase. The topological spin of $s$ is $i$. The only nontrivial (not $1$) components of the $F$ and $R$ symbols are $\left[F^{sss}_s\right]_{\mathbb{1},\mathbb{1}} = -1$ and $R^{ss}_{\mathbb{1}}=i$.

A $Z_2$ topological order contains three types of nontrivial anyons: the bosonic gauge charge $e$, the bosonic gauge flux $m$ and their composite $f=e\times m$. All three anyons are abelian. The fusion rules are given by
\be
e\times e = m\times m = f\times f = \mathbb{1}, \ e\times m = f, \ m\times f = e, \ f\times e = m
\ee
where again $\mathbb{1}$ denotes the vacuum sector of the topological phase. The topological spins $\theta_a$ and braiding statistics $\theta_{ab}$ are given by
\be
\theta_e=\theta_m=1, \theta_f=-1, \theta_{em}=\theta_{mf}=\theta_{fe}=-1
\ee
The $F$ symbol of the $Z_2$ topological order can be totally trivial (all $1$) while the $R$ symbols can take the form
\be
R^{ee}_{\mathbb{1}} = R^{mm}_{\mathbb{1}} = 1, R^{ff}_{\mathbb{1}} = -1, R^{em}_{f} = R^{ef}_m = R^{fm}_e = 1, R^{me}_f=R^{fe}_m = R^{mf}_e = -1.    
\ee

\section*{Appendix B: Symmetry Protected Topological Order and Their Corresponding Gauge Theories}

Symmetry protected topological (SPT) order exists in gapped systems with global symmetry. The ground state of the system does not spontaneously break the symmetry, has no fractional excitation, yet cannot be smoothly connected to a product state without explicitly breaking the symmetry. The nontrivial natural of the SPT order can be manifested in two ways: 
\begin{enumerate}
\item
As nontrivial edge states which must either be gapless, spontaneously break symmetry or support anomalous SF pattern (with bulk dimension $\geq$ 3) as long as the global symmetry is not explicitly broken.
\item
For SPT phases with unitary on-site symmetry, gauging the symmetry results in nontrivial gauge theories whose gauge fluxes have nontrivial braiding statistics. 
\end{enumerate}

It was shown that a large class of SPT phases in boson / spin systems in dimension $d$ has a one to one correspondence with equivalence classes of group cocycles $H^{d+1}(G,U(1))$\cite{Chen2012,Chen2013a}. For unitary $G$, gauging the symmetry results in $d$ dimensional Dijkgraaf-Witten (DW) gauge theory characterized also by the equivalence class of cocycles\cite{Dijkgraaf1990}. For example, the double semion theory in $2D$ is a DW gauge theory of gauge group $Z_2$ corresponding to the nontrivial element in $H^3(Z_2,U(1))$. For definition of group cocycles and discussion of their relation with SPTs and gauge theories , see \Ref{Chen2012,Chen2013a, Dijkgraaf1990}. 

For the discussion in this review, it suffices to know that the set of equivalence classes of group cocycles in $H^{d+1}(G,U(1))$ forms an abelian group. For $d=0$, elements in $H^{1}(G,U(1))$ corresponds to one dimensional representations of $G$ which have a one to one correspondence with symmetry charges of $G$. For $d=1$, elements in $H^2(G,U(1))$ corresponds to projective representations of $G$ with $U(1)$ coefficient; such projective representations are realized as degenerate edge states of one dimensional SPT phases. We collect a few simple mathematical results of $H^{d+1}(G,U(1))$, for $d\geq 1$.
\begin{align*}
H^2(SO(3),U(1))=Z_2,  \ H^2(Z_2\times Z_2,U(1)) = Z_2,  \ H^2(Z_2^T,U(1))=Z_2,   \ H^3(Z_n,U(1))=Z_n \\
H^3(Z_2^T,U(1))=Z_1,  \ H^4(Z_n,U(1))=Z_1, \ H^4(Z_n\times Z_n,U(1)) = Z_n\times Z_n,  \ H^4(Z_2^T,U(1))=Z_2
\end{align*}
Here $Z_n$ denotes the cyclic group of order $n$, $Z_2^T$ denotes the $Z_2$ group of time reversal symmetry.

Correspondingly some simple examples of SPT phases include: 
\begin{enumerate}
\item Haldane phase in 1D with spin rotation symmetry\cite{Haldane1983a}; the edge spin $1/2$ must be degenerate unless spin rotation symmetry is broken.
\item Topological insulators and superconductors in 2D with time reversal symmetry; the edge carries gapless helical modes\cite{Kane2005}.
\item Bosonic SPT with $Z_2$ symmetry in 2D\cite{Chen2011}; gauging the $Z_2$ symmetry results in the double semion topological order\cite{Levin2012}.
\item Topological insulators in 3D with time reversal symmetry\cite{Fu2007,Moore2007,Roy2009}; the 2D boundary carries a single Dirac cone.
\end{enumerate}

\end{document}